\documentclass[twocolumn,showpacs,amsmath,amssymb,pre]{revtex4-1}

\newcommand{\bes}{\begin{subequations}\bea}
\newcommand{\ees}{\eea\end{subequations}}
\newcommand{\be}{\begin{equation}}
\newcommand{\ee}{\end{equation}}
\newcommand{\bea}{\begin{eqnarray}}
\newcommand{\ba}{\begin{array}}
\newcommand{\eea}{\end{eqnarray}}
\newcommand{\ea}{\end{array}}

\begin{document}

\preprint{APS/123-QED}

\title{Diffusion in nonuniform temperature and its geometric analog}

\author{Matteo Polettini} 
 \email{matteo.polettini@uni.lu}
\affiliation{Complex Systems and Statistical Mechanics, University of Luxembourg, Campus
Limpertsberg, 162a avenue de la Fa\"iencerie, L-1511 Luxembourg (G. D. Luxembourg)}

\date{\today}

\begin{abstract}
We propose a Langevin equation for systems in an environment with nonuniform temperature. 
At odds with an older proposal, ours admits a locally Maxwellian steady state, local equipartition holds and for detailed-balanced (reversible) systems statistical and physical entropies coincide. We describe its thermodynamics, which entails a generalized version of the First Law and Clausius's characterization of reversibility. Finally, we show that a Brownian particle constrained into a smooth curve behaves according to our equation, as if experiencing nonuniform temperature. 
\end{abstract} 



\maketitle

It is well known that noise drives particles along the temperature gradient, giving rise to thermophoresis, or thermodiffusion. ``Pebbles in a driveway accumulate on the side'', wrote Landauer suggestively. ``In the driveway they are agitated (hot region). They are left undisturbed on the side (cold region).''  \cite{landauer} He had long sharpened his intuition on the problem \cite{landauer2}, and came to argue that, in the light of the equipartition theorem for the root mean square velocity $ \bar{v} \sim \sqrt{T}$, detailed balancing of the currents $p_h \bar{v}_h$ and $p_c \bar{v}_c$, respectively out of a hot region and of a cold one, implies a spatial density $p \sim 1/\sqrt{T}$. He admitted though  that this all too simple argument did not do justice to more complicated situations where the geometry of the mean free path changes with temperature, and that other proposals were valuable, in particular Van Kampen's $p \sim 1/T$, which followed from the analysis of a Langevin equation with state-dependent diffusion coefficient \cite{vankampenIBM}. Since then, Van Kampen's theory has been used by various authors to discuss the stochastic energetics of nonuniform temperature systems \cite{matsuo} (see also Sekimoto's monograph \cite[\S 1.3.2.1, \S 4.1.2.2]{sekimotobook}), to derive the Soret coefficient \cite{bringuier}, and more recently to find an anomaly between overdamped thermodynamics and thermodynamics with finite, but small, inertia \cite{celani}. 

However, when applied to reversible (i.e.  detailed balanced) systems, Van Kampen's theory displays some odd peculiarities. The steady state is not locally Maxwellian (local thermal equilibrium). Equipartition fails: Temperature is not the average kinetic energy. The conventional definition of heat does not allow to match statistical and physical notions of entropy, a cornerstone principle of equilibrium statistical mechanics. Moreover, generalization to $n$-dimensional systems with non-diagonal diffusion matrix leads to nonequilibrium steady states even in the absence of external forces \cite{polettini}. All these effects, if a bit controversial from a foundational perspective, might well occur and are of great interest; after all equipartition and Gaussianity should be ascribed to ideal situations. However, so far we lack a basic ideal model.

\subsubsection{Scope and overview}

In this work we enforce the principle of local thermal equilibrium to obtain a different theory that marries well with the thermodynamics of nonuniform temperature systems and panders to Landauer's intuitions. We  first derive the relevant equations and discuss their dynamical properties and time scales. We then define proper equilibrium thermodynamic potentials, discuss conditions for the occurrence of equilibrium and dwell into thermodiffusion in nonequilibrium regimes. We already pinpoint that boundary conditions on the solutions to our equations allow a steady state that satisfies detailed balance if and only if local thermal equilibrium holds.  We derive some peculiar experimentally accessible consequences along the way.

At last, before drawing conclusions, we introduce the theme of a Brownian constrained particle. It was Maupertuis who first realized that a spatially dependent kinetic energy could be interpreted as a deformation of the metric of space. Along this line, we prove that the same equation that describes diffusion in nonuniform temperature also describes the motion of a Brownian particle in $m$-dimensional space at uniform temperature $T_0$, which is constrained into a smooth curve. The equivalence between thermodynamics and geometry suggests an interpretation of inverse temperature as a metric
\be
T(x) = T_0 / t(x)^2, \label{eq:templen}
\ee
where $t(x)$ is the adimensional norm of the tangent vector to the curve and $x$ is a curve parameter, with units of a spatial variable. In particular, the theory is covariant: A suitable choice of coordinates can modify the temperature profile, making it locally uniform. As a further connection between geometry and thermodynamics, the temperature profile can also be made globally uniform by a coordinate transformation if and only if the system satisfies detailed balance. Generalization to $n$ dimensions is deferred to a parallel publication \cite{polettini}, where the geometrical aspects are explored in greater depth.

Before plunging into details, let us first take a glance at a more physical byproduct of the theory, viz. the following generalization of the first law of thermodynamics for systems that experience a varying environmental temperature ($k_B$ set to unity):
\be
\delta Q = dU - \delta W +  \big(n- U/T\big) dT. \label{eq:first}
\ee
The one-dimensional case will be given a precise formulation in the formalism of stochastic thermodynamics. Equation (\ref{eq:first}) is a straightforward generalization for $n$ independent degrees of freedom. It must be emphasized that both our theory and Van Kampen's imply modified first laws, whose implications have yet to be  fully explored, to our knowledge. Maybe the most remarkable fact is that, even keeping the internal energy fixed and performing no mechanical work, yet heat flows when the system meets a temperature variation. Equation (\ref{eq:first}) is in agreement with the usual first law once the last term is identified as a ``thermal work'' $- \delta W_{th}$ that needs be performed on the system to help it climb up against the temperature gradient, or that can be extracted when the system follows its natural tendency towards colder temperatures. For an overdamped system, performing a transition on the ``equipartition shell'' $U = nT/2$, the thermal work performed reads $- \delta W_{th} = ndT/2 = dU$. Along an adiabatic transformation the total mechanical work that needs be performed on the system is twice the kinetic energy $-\delta W= 2 dU$.

\subsubsection{Dynamics} Let us make all of this more precise. In the following $\zeta_t$ is white Brownian noise, with $\langle \zeta_t \zeta_{t'} \rangle = \delta(t-t')$. We use the Stratonovich convention on the stochastic differential, which allows us to perform standard differential calculus. For the sake of simplicity we omit the notation ``$\,\circ \,$'' for the stochastic differential. We contract  the  spatial derivative with $\partial$ and the derivative with respect to velocities with $\partial_v$. Degrees of freedom have unit inertia; think of a Brownian particle of mass $m = 1$ for definitiveness. Below, dimensional units can be recovered by scaling $T \to k_B T/m$, $V \to V/m$.  We first suppose that $x$ either belongs to a cyclic domain or to the real line (periodic boundary conditions), with temperature and forces smooth functions; later on we will comment on discontinuities and domains with boundaries. 

Conventional wisdom supports the following Langevin equation for a particle in a bath at uniform temperature
\be
\dot{x}_t = v_t, \quad \dot{v}_t = -  \partial V(x_t) - \gamma v_t + \sqrt{2\gamma  T} \, \zeta_t, \label{eq:temp}
\ee
whose corresponding diffusion (Kramers) equation suitably affords the Maxwell-Boltzmann distribution
$P^\ast \propto \exp - T^{-1} \left( v^2/ 2  + V \right)$ as its steady state. Generalization to media with state-dependent temperature is legitimately carried out by many authors by replacing $T$ by $T(x)$ in Eq.(\ref{eq:temp}) \cite{vankampenIBM,matsuo,sekimotobook,bringuier,celani}, on the basis that locally, where temperature is approximately uniform, it returns the original equation. This is also true of any equation that contains derivatives of the temperature. Each such theory has its peculiar properties and leads to distinctive predictions. In particular, the former choice leads to a discrepancy between $T(x)$ and the notion of temperature as average kinetic energy, as we argue by considering the corresponding Kramers generator with state-dependent temperature:
\be
\hat{L}_{\mathrm{V.K.}} \, P = - v \, \partial P +  \partial_v \left[(\partial V + \gamma v) P +   \gamma T  \partial_v P  \right].
\ee
The steady distribution in the vicinity of a given fixed point $x$ turns out not to be locally Maxwellian --- nor easily computable --- as one can appreciate by plugging the ansatz  $R = r(x) \exp - k(x) v^2$. For no choices of $k(x), r(x)$ does in fact $\hat{L}_{\mathrm{V.K.}} R = 0$. It follows that the average kinetic energy of a particle at $x$ does not coincide with $T(x)/2$. In other words,  local equipartition fails. The steady solution only boils down to the locally Maxwellian form in the overdamping limit $\gamma \to \infty$, while corrections to local equipartition of order  $\gamma^{-1}$ can be shown to also contain derivatives of the temperature \cite[Supplementary Material]{celani}. This poses an interpretational problem: In which sense is $T(x)$ ``temperature''? When the system is a binary mixture, temperature can indeed be defined as the average kinetic energy of the environmental component. Although, very often the physical nature of noise is not discernible, in which case temperature might be self-consistently defined as average kinetic energy of the system's degrees of freedom. In these cases Van Kampen's theory is not applicable in an obvious way.

Alternatively, we can use the same \textit{a posteriori} requirement on the form of the desired Langevin equation as for Eq.(\ref{eq:temp}), by postulating that the steady state is
\be
P^\ast(v,x) \propto \exp - \left[ v^2/ 2T(x)  + \psi(x)\right].\label{eq:nut}
\ee
It seems reasonable, in fact, that locally in a small region nearby $x$ where temperature is approximately uniform, the velocity distribution is again Maxwellian and equipartition holds, regardless of how temperature varies outside that small region. This  assumption can in principle be subjected to experimental validation, although it might be complicated to control the statistics of particles' velocities sufficiently close to a given point. We expect that this assumption should hold whenever that small neighborhood is nevertheless big enough to enable the Brownian particle to collide with particles in the underlying medium, all the same temperature, a sufficient number of times. In other words the mean free path $\gamma^{-1} \sqrt{k_B T/m}$ shall be much smaller than the scale of variation of the temperature $T/\partial T$. 

We plug Eq.(\ref{eq:nut}) into a generic Kramers-type generator
\be
\hat{L}_{\mathrm{th}} \, P^\ast = - v\, \partial   P^\ast+ \partial_v \left( D  \partial_v P^\ast   -  f P^\ast   \right) = 0 \label{eq:lannut},
\ee
where we allowed for a state-dependent diffusion coefficient $D(x)$, and $f(x,v)$ is the drift. We obtain:
\be
- v^3 \partial T /2  + v T^2 \, \partial \psi   - D T  + D  v^2 -  T^2 \partial_v f + T f v = 0. \label{eq:vanish}
\ee
For this expression to vanish, the latter drift terms must necessarily counterbalance all others, order-by-order in the velocities. Hence the drift must have the form
\be
f(x,v) = f_0(x) + f_1(x) v + f_2(x) v^2. 
\ee
Plugging in Eq.(\ref{eq:vanish}), after some work we obtain
\be
f_0 =  \partial T - T \partial \psi, \quad
f_1 = - D/T, \quad
f_2 = \partial \ln \sqrt{T}. \label{eq:relations}
\ee
The second expression is a generalized version of Einstein's relation. A convenient choice for the diffusion coefficient, making us closer to Eq.(\ref{eq:temp}), is $D = \gamma T$. It must be emphasized though that many other equally valuable models yield the same steady state. The third term gives a  sort of Rayleigh  drag force, which is square in the velocities and depends on the temperature gradient. Finally we obtain the following Kramers equation for a generic probability density $P=P(x,v ,t)$
\be
\frac{\partial P}{\partial t} + v\, \frac{\partial P}{\partial x} 
+ \left(f_0 + \frac{v^2}{2T} \frac{\partial T}{\partial x}  \right) \frac{\partial P}{\partial v}
= \gamma \frac{\partial}{\partial v} \left(  T  \frac{\partial P}{\partial v}  + vP\right)\label{eq:kramers},
\ee
and its corresponding Langevin equation
\be
\dot{v}_t = f_0(x_t) - \gamma v_t +  v_t^2 \, \partial \ln \sqrt{T(x_t)} + \sqrt{2 \gamma T(x_t)} \, \zeta_t , \label{eq:lantem}
\ee
where $f_0$ plays the role of driving force.  The left-hand side of the Kramers equation can be cast as $(\partial_t + \hat{L})P$, where $\hat{L}$ is the Liouvillian
\be
\hat{L} P = \frac{\partial H(x,\pi)}{\partial \pi} \frac{\partial P(x,v(x,\pi))}{\partial x} - \frac{\partial H(x,\pi)}{\partial x}  \frac{\partial P(x,v(x,\pi))}{\partial \pi},
\ee
with Hamiltonian $H = T(x)\pi^2/2 + \Phi$, momentum $\pi = v/T(x)$, and $\Phi$ the potential of $f_0/T = - \partial \Phi$. The effective inertia of the particle $\pi/v$ depends on temperature, being smaller where temperature is higher. The steady probability measure  in position/momentum space is the Gibbs measure $P^{\ast}dx \, dv = \exp-H(x,\pi) dx \, d\pi$. 

Equations (\ref{eq:kramers} and \ref{eq:lantem}) are the central equations of this work. We will assume throughout that they locally describe the interaction of a Brownian particle in contact with a medium with varying temperature. The occurrence of the locally Maxwellian steady state depends on smooth boundary conditions. In general, different boundary conditions yield different solutions accordingly.

In the overdamping limit where $\gamma$ is very large, when relaxation to a locally Maxwellian state long precedes spatial relaxation, the Kramers generator reduces to a Fokker-Planck generator for the spatial distribution $p = \int dv \,P$. While Van Kampen's theory leads to the It\=o generator \cite{celani}, ours yields the Stratonovich generator
\be
\partial_t p =  - \gamma^{-1} \partial \left[f_0 \, p - \sqrt{T} \, \partial \left( \sqrt{T} p \right) \right] = - \partial J, \label{eq:firstorder}
\ee
where $J$ is the probability current. The latter diffusion equation corresponds to the first-order stochastic differential equation $\dot{x}_t =  f_0(x_t)/\gamma + \sqrt{2 T(x_t)/\gamma}\, \zeta_t$ in Stratonovich convention. For vanishing forces, the average drift out of a given state $x$ is to the hot,
\be
\langle dx_t \rangle_x = \langle \sqrt{2 T(x_t)/\gamma}\, \zeta_t \rangle_x \, dt = \frac{\partial T(x)}{2\gamma} dt, 
\ee
which marks a difference with respect to the null average increment in Van Kampen's theory \cite{sekimotobook}. Almost paradoxically, at an equilibrium steady state particles drift aside even in the absence of a steady flux, a well-known phenomenon in state-dependent diffusion  \cite{lancon}. 

On the real line, the heat kernel
\be
K(x,x_0,t) \propto \frac{1}{\sqrt{T(x) t } } \exp - \frac{\gamma}{4t} \left( \int^x_{x_0} \frac{dx'}{\sqrt{T(x')}} \right)^2
\ee
solves Eq.(\ref{eq:firstorder}) with null forces and vanishing boundary conditions at infinity, given the point-wise initial condition $\delta(x-x_0)$ \cite{fa}. Sufficiently close to the initial state $x = x_0 + \delta x$, at small times
\be
K(x_0 + \delta x,x_0,t) \approx \frac{1}{\sqrt{T(x_0)t} } \exp - \frac{\gamma \left(x - x_0 \right)^2}{4t T(x_0)}  \label{eq:hkapprox}
\ee
to second order. Then $T(x_0)^{-1}$ affects the typical time for spatial relaxation near a point: The hotter, the faster. At larger distances and long times there will be corrections due to curvature (as described in Ref. \cite{castro}), but basically spatial relaxation is ruled by the value of the temperature and not by its variation, as far as this is not too sudden. In the typical time $\gamma^{-1}$ where the system attains a locally Maxwellian state, the mean square displacement around $x_0$ is $\delta x \sim \sqrt{T}/\gamma$. Since we assumed $\gamma \gg \partial T / \sqrt{T}$, we find that $\delta x \ll \partial T / T$, i.e., the region where the approximated Eq.(\ref{eq:hkapprox}) holds is smaller than the typical scale where $T$ varies.

The overdamping limit is carried out in detail in Appendix \ref{overdamping}. Notice that, at stake with the uniform temperature case, the overdamped stochastic differential equation cannot be obtained by just neglecting the acceleration in Eq.(\ref{eq:lantem}).

\subsubsection{Heat balance at equilibrium} In this and the following sections we give some physical insights, revisiting Landauer's and Buttiker's works \cite{landauer,buttiker} in the light of classical and stochastic thermodynamics \cite{sekimoto,matsuo,esposito}.

We first consider the case where the boundary conditions on Eq.(\ref{eq:lantem}) allow Eq.(\ref{eq:nut}) as the equilibrium  state. The first relation in Eq.(\ref{eq:relations}) can be integrated to give 
\be
\psi(x) = - \int^x \frac{\delta W}{T} + \ln T(x)
\ee
where $\delta W =   f_0(x) dx$  is  the work exerted on the system along displacement $dx$. 
The steady state now reads
\be
P^\ast \propto \frac{1}{T} \exp - \left( \frac{ v^2}{2 T } -
\int \frac{\delta W}{T} \right) = \exp - S(x,v), \label{eq:ss}
\ee
where $S$ is an information-theoretic generalization of Boltzmann's entropy function of state $(x,v)$ with respect to the steady state \cite{farewell} (for a motivation from large deviation theory, see Ref.\cite{touchette}). When evaluated along a fluctuating trajectory in phase space (``on shell''), it is also known as self-information at the steady state. It serves as definition of entropy of state $(x_t,v_t)$ along a stochastic trajectory, at late times \cite{seifert}, while at shorter times one shall employ $S(x,v,t) = - \ln P(x,v,t)$.   

Performing the Gaussian integral in Eq.(\ref{eq:ss}) with respect to velocities we obtain the spatial density
\be
p^\ast(x) \propto \frac{1}{\sqrt{T(x)}} \exp \int^x  \frac{\delta W}{T}. \label{eq:sss}
\ee
When external forces vanish, $f_0 = 0$, the steady spatial distribution is $p \propto 1/\sqrt{T}$, reconciling to Landauer's intuitive picture. In particular,  Eq.(\ref{eq:sss}) makes the current in the overdamped equation vanish, so that local thermal equilibrium implies (global) equilibrium. If instead we were to impose a uniform pressure $pT = \mathrm{const.}$, we would obtain a current profile $J = \partial \ln \sqrt{T}$ from the cold to the hot. Analogous conclusions were reached by Balakrishnan and Van den Broeck in their study of the Jepsen gas, where one particle (the piston) is singled out between two one-dimensional gases of identical particles interacting solely by elastic collisions, and kept at different temperatures \cite{balak}. In that case it is the gas's density that settles into the Landauer state $1/\sqrt{T}$. In our case, it is tempting to figure out Brownian particles sitting on top of the gas as they arrange themselves according to the same spatial distribution as the underlying medium. 

The notion of entropy might be associated to that of {\it internal heat production} 
\be
T dS = \delta Q^{in} ~=~ \stackrel{\delta Q^{en} }{\overbrace{d\left(v^2/2 \right) - \delta W }} ~+~ \stackrel{\delta Q^{th}}{\overbrace{dT - v^2 d \ln \sqrt{T}}}, \label{eq:defheat}
\ee
which when evaluated on shell yields the amount of heat that is produced when the system performs a transition from $(v,x)$ to $(v+dv,x+dx)$.  Overbraces are used to define the \textit{energetic} and the \textit{thermal} heat productions. The latter vanishes when temperature is uniform, the former accounts for the kinetic energy variation and for the work done by the external force. The first term contributing to the thermal heat production can be interpreted in the light of Fourier's first law of heat conduction, and it was accounted for by Landauer. The second term is peculiar of our approach; it corresponds to the new drag force.

The internal heat production is counter-balanced by an {\it external heat flux} towards the environment
\be
\delta Q^{ex}_t 
=  - \left[ \gamma v_t  - \sqrt{2 \gamma T(x_t)} \, \zeta_t \, \right] dx_t +  dT(x_t),  \label{eq:heatonshell2}
\ee
in all analogous to Sekimoto's definition in uniform temperature, but for the last term $dT$. The total heat flux at a steady state is given by $\delta Q^{in} -  \delta Q^{ex}$. Plugging Eq.(\ref{eq:lantem}) into Eq.(\ref{eq:defheat}) we find that the two contributions coincide and the total heat flux vanishes, as one expects at an equilibrium steady state. {\it A posteriori} this is the main reason for introducing the above definition, and we deem it a stength point of our approach. Exporting Sekimoto's definition into Van Kampen's theory, as done in Ref. \cite[Eq.(40)]{matsuo}, due to the fact that the internal entropy $- \ln P^\ast$ will generally contain higher-order terms in the velocity, and Sekimoto's definition only contains the second-order kinetic term, the internal heat production will not balance external heat flux, regardless of corrections $\propto \partial T$, thus yielding a non-null steady entropy production along a trajectory even when the corresponding overdamped Fokker-Planck equation displays a null steady current. This can be regarded as yet another perspective on the anomalous behavior of Van Kampen's theory, discussed by Celani {\it et al.} \cite{celani}, which our theory does not display.

\subsubsection{Generalized first law and thermodynamic potentials}

Letting $\delta Q = \delta Q^{in} = \delta Q^{ex}$, the generalized first law Eq.(\ref{eq:first}) follows from Eq.(\ref{eq:defheat}) once the internal energy $U =  v^2/2$ is identified. We now dwell on its interpretation. The last term  $U/T d T$  is not at all unusual. In fact, from $\delta Q = T dS = d(TS) -  S d T$ and from the definition of free energy $\mathcal{F} =  \mathcal{U}- TS$ we  have
\be
\delta  Q = d( \mathcal{U}-\mathcal{F}) - \frac{ \mathcal{U}-\mathcal{F}}{T} dT, \label{eq:prefirst}
\ee
wherefrom our first law will follow, after a sensible identification of the subsidiary thermodynamic potentials  $\mathcal{U}$ and $\mathcal{F}$ is done. To accomplish this we split the process into two subprocesses. Consider the conditional probability $P^\ast(v,x)/p^\ast(x)$ of having velocity $v$ at a given state $x$, and define the conditional entropy $S_1 = - \ln (P^\ast/p^\ast)$ and the spatial entropy $S_2 = - \ln p^\ast$. The total entropy obviously splits into the two contributions, $S = S_1 + S_2$. We define the conditional energy at a point as $U_1 = U$. Hence by definition of free energy $F_{i} = U_{i} - T S_{i} $ we have 
\be
F_1 (x) = - \frac{T(x)}{2} \ln T(x).
\ee
Notice that, not depending on $v$, at fixed $x$ the free energy is a constant. Its $x$-dependence suggests to take $F_1$ also as definition of the spatial free energy $F_2 = F_1 = F$, so that the spatial energy turns out to be
\be
U_2(x) = - T(x) \int^x \frac{\delta W}{T}.
\ee
With these definitions we obtain the two sub-laws
\begin{subequations} \label{eq:split}
\bea
\delta Q_{1} & = & d U + dT/2 - U/T \,dT   \\
\delta Q_{2} & = & - \delta W + dT/2   \label{eq:second}
\ees
which combined together return the above heat flux $\delta Q= \delta Q_1 + \delta Q_2$, given $\mathcal{U}= U_1 + U_2$ and  $\mathcal{F} = 2F$. Consider Eqs.(\ref{eq:split}): The second is the heat balance equation in first-order formalism, while the first provides a further contribution, which is peculiar to second-order theory. The average kinetic energy at a point $x$ decreases exponentially to its equiparted value $T(x)/2$, as standard manipulations in stochastic calculus yield 
\be
\frac{d}{dt} \left\langle v^2\right\rangle_x = - 4\gamma \left[\left\langle v^2\right\rangle_x  -  T(x) \right],
\ee
where the average is conditioned to $x$ at time $t$. In the overdamping limit, the relic of the internal energy $dT/2$ survives in Eq.(\ref{eq:second}), while on average the first equation reduces to $\delta Q_{1} = d T/2$; hence reduction from phase space to state space removes one equiparted degree of freedom from the definition of heat.

\subsubsection{\label{noneq} Nonequilibrium thermodynamics}

According to classical thermodynamics, equilibrium is characterized by $\delta Q^{ex}/T$ being an exact differential. In other words there shall exist a state function $S_C$ --- the Clausius entropy --- such that $dS_C = \delta Q^{ex}/T$ in all of its domain (condition of detailed balance, or reversibility). A peculiarity of our definition of external heat flux is that, when detailed balance holds, Clausius's entropy exists and it coincides with Boltzmann's. However, while the latter  is always well defined as minus the logarithm of the steady probability density, the former might not. When the temperature profile and the steady probability density are smooth functions, Clausius's entropy exists if and only if Clausius's criterion for reversibility holds 
\be
\oint_{(x,v)}  \frac{\delta Q^{ex}}{T} = - \oint_{x}  \frac{\delta W}{T}  = 0, \label{eq:loop}
\ee
where the two integrals are taken along closed loops respectively in phase space and in state space. The equivalence of these two integrals is straightforward given that all terms involving velocities are exact differentials in the expression for the heat, as far as temperature is a smooth function. The condition in Eq.(\ref{eq:loop}) becomes more pertinent in higher dimensions, where it holds when $\vec{f}_0 = - T \nabla \Phi$, with $\Phi$ a smooth potential. In one dimension a violation of Eq.(\ref{eq:loop}) can only occur as a topological effect, due to a discontinuity of the external force.

However, even a discontinuous force might lead to equilibrium steady states, if the probability distribution and/or the temperature have discontinuities themselves that counter-balance those of the force. This depends on the boundary conditions on the probability. We only mention, without further discussion, that a more modern and complete condition for equilibrium in terms of vanishing macroscopic affinities \cite{schnak,gauge} requires us to also keep into account discontinuities of the function $ T P^\ast$ as an information-theoretic contribution to dissipation. Rather than a full treatment, we prefer to give an example in the overdamping limit to better convey these aspects and derive one experimentally accessible result.

Consider an overdamped Brownian particle in a closed interval whose extremities are put in contact with heat reservoirs at different temperatures $T_c$ and $T_h$. No external forces are exerted. Equation (\ref{eq:firstorder}) allows infinitely many steady solutions, for any value of the uniform steady current $J^\ast$. Integrating $\partial (\sqrt{T} \, p^\ast ) = - J^\ast/\sqrt{T}$ we obtain
\be
J^\ast  \int_c^h   T^{-1/2}  = p^\ast_c \, \sqrt{T_c}  - p^\ast_h  \, \sqrt{T_h} ,  \label{eq:prediction}
\ee
which establishes a relationship between the steady current and the boundary values of the probability distribution. In Van Kampen's theory, the analogous result to Eq.(\ref{eq:prediction}) is
\be
(x_h - x_c) J^\ast  = p^\ast_c \, T_c  - p^\ast_h  \, T_h.
\ee
In ours, the Landauer steady state that we initially enforced is the one that satisfies detailed balance, $p^\ast_c \, \sqrt{T_c} = p^\ast_h  \, \sqrt{T_h}$. By means of particle reservoirs, an external agent can control the particle densities at the extremities and  impede relaxation, triggering directed particle transport. The steady probability current is given by the average velocity of particles by $J^\ast = \int  P^\ast(x,v)v \, dv$. In order to have $J^\ast \neq 0$, the steady state $P^\ast(x,v)$ shall not be even in the velocities, and in particular it can't be Gaussian. Local thermal equilibrium then fails, compatibly with well-known results from specific models \cite{dhar}. This discussion reveals that, within our theory (specified by the equations), the steady state of a specific model (a solution to our equations) obeys local thermal equilibrium if and only if it is globally equilibrated. 

\subsubsection{Constrained Brownian particle} Let us now turn page. We consider a Brownian particle living in $n$-dimensional Euclidean space with uniform temperature $T_0$. Extrinsic coordinates and velocities are called $(z^i,u^i)$, with spatial derivatives $\partial_i = \partial/\partial z^i$. Einstein's convention on index contraction is employed.

At first, the particle is assumed to obey the standard $n$-dimensional  Langevin equation
\be
\dot{u}^i = f^i - \gamma u^i + \sqrt{2\gamma T_0} \,  \zeta^i =: F^i,
\ee
and  then forced into a smooth curve with the introduction of $n-1$ independent smooth holonomic constraints $\Phi^a(z_t) = 0$. Taking the first and second time derivatives we obtain the following constraints on the velocity and the acceleration,
\be
\Sigma_i^a u^i = 0, \qquad \Sigma_i^a  \dot{u}^i + \Gamma^a_{ij}  u^i u^j = 0 , \label{eq:constraint}
\ee
where we defined
\be
\Sigma_i^a = 	\partial_i \Phi^a, \qquad \Gamma^a_{ij} = \partial_i \partial_j \Phi^a.
\ee
We assume that the matrix $\Sigma_i^a$ has rank $n-1$ and further define $G^{ab} =  \Sigma_i^a  \Sigma_j^b  \delta^{ij}$, which is positive definite.

The key step is to appeal to Gauss's principle of least constraint, 
which states that on-shell accelerations minimize the so-called Gauss curvature
\be \mathcal{C}(\dot{u}) =  \delta_{ij} ( \dot{u}^i - F^i) ( \dot{u}^j - F^j)/2\ee
compatibly with the constraints. Constraints are taken into account by introducing Lagrange multipliers $\lambda_a$. Setting the first variation of Gauss's curvature with constraints to zero,
\be
\frac{\delta}{\delta \dot{u}^k}\left[ \mathcal{C}(\dot{u}) + \lambda_a \left( \Sigma_i^a  \dot{u}^I + \Gamma^a_{ij}  u^i u^j \right)\right] = 0,
\ee
we obtain $\dot{u}^i =  F^i -  \lambda_a   \delta^{ij}    e_j^a$.
Lagrange multipliers can be determined by plugging this solution into the constraint equation Eq.(\ref{eq:constraint}), yielding
\be \lambda_a =  G_{ab}  (  \Gamma^b_{ij}  u^i u^j  + \Sigma_i^b F^i). \ee
We obtain a Langevin equation in extrinsic coordinates
\be
\dot{u}^i = -  \delta^{ij}   \Sigma_j^a  G_{ab} \Gamma^b_{kl} u^k u^l  + (\delta^i_j - G_{ab}     \Sigma_k^a  \Sigma_j^b \delta^{ki}) F^j,
\label{eq:extrinsic}
\ee
which will now be expressed in intrinsic coordinates. We parametrize the curve $x \to z^i(x)$, and introduce the tangent vector  $t^i=  \partial z^i$ with norm
\be
t(x) = \sqrt{\delta_{ij} t^i(x) t^j(x)}.
\ee 
The intrinsic velocity $v_t = \dot{x}_t$  can be expressed in terms of the extrinsic velocity by taking the time derivative of $z^i_t = z^i(x_t)$, which yields $u^i_t  =  v_t  t^i(x_t)$. Hence:
\be
v_t = t(x_t)^{-1}  \sqrt{\delta_{ij} u^i_t u^j_t }.
\ee
Finally, we take the stochastic differential of  this expression and plug Eq.(\ref{eq:extrinsic}) in. After some standard calculus:
\bea
\dot{v} & = &   - v^2 \, \partial \ln t + t^{-2} \left(\delta_{ij}  - G_{ab}      \Sigma_k^a  \Sigma_j^b \delta^{k}_i\right) t^i  F^j  \label{eq:covariant} \\
& = & t^{-2} \delta_{ij} t^i f^j - \gamma v -  v^2 \, \partial \ln t + \sqrt{2\gamma T_0} \, t^{-2} \delta_{ij}\,  t^i \zeta^j,\nonumber 
\eea
where we repeatedly used the constraint $\Sigma^a_i u^i = 0$. Notice that 
$\zeta_t = t^{-1} \delta_{ij} t^i  \zeta^j_t$ is Brownian white noise. It is easily proven in fact that it has null average, it is $\delta$-correlated with unit diffusion coefficient, and it is normally distributed as the sum of independent normal stochastic variables have normal distribution. Therefore, identifying the force $f_0 = t^{-2} \delta_{ij} t^i f^j$ and, most importantly, the temperature as square the inverse length as in Eq.(\ref{eq:templen}), we re-obtain Eq.(\ref{eq:lantem}), which can now be interpreted geometrically  as a noisy damped geodesic equation, with the term involving Christoffel coefficients playing the role of the Rayleigh-type drag force. If the external force is conservative, $f^j = - \delta^{ij} \partial_j \Phi$, the driving force $f_0 = - T \partial \Phi$, while possibly non-conservative, satisfies detailed balance. As a curiosity, notice that in the expression for the square modulus of the extrinsic acceleration of the particle  (setting $T_0 =1$)
\be
a^2 =  \delta_{ij} \dot{u}^i \dot{u}^j  = T^{-1} \left(  \dot{v}^2 - \dot{v}\, U \partial \ln T  + 4\kappa^2 T U^2 \right),
\ee
there appears the intrinsic acceleration $\dot{v}$ and correcting terms, the second of which contains the squared Gauss curvature $\kappa^2 = \delta_{ij}\partial t^i \partial t^j$, let $U=v^2/2$.

Parametrizing the curve with the length itself, in such a way that the tangent vector is always normalized, $\delta_{ij} t^i t^j = 1$, we obtain the standard Langevin equation with uniform temperature: Temperature can be made (locally) uniform by a (local) coordinate transformation.  This is the tip of the iceberg: beneath lies the general covariance of the theory under coordinate and gauge transformations \cite{polettini}.

\subsubsection{Conclusions} In order to blend geometry and thermodynamics, consider as an example Brownian particles floating in a fluid contained in a torus-shaped recipient, with minor radius much smaller than the major radius. Suppose the fluid is heated in a sector of the torus. According to our theory, Brownian particles will arrange themselves in an equilibrium steady state that displays local thermal equilibrium. Their motion is equivalent to that of particles in a deformed torus, with higher temperature where the toroidal shape contracts.  Eq.(\ref{eq:templen}) says that temperature can be interpreted as a measure of effective length: Random transitions are more probable to shorter distances, as if due to hotter temperatures. This picture marries well with Smerlak's proposal of ``tailoring'' the diffusion of light in suitably engineered metamaterials, with diffusion coefficients acting as an effective curvature of space (-time) \cite{smerlak}. Several steps in our work also resound with the techniques used by Zulkowski and coworkers \cite{zulkowski} to build geodesics in the space of thermodynamic parameters out of the inverse diffusion tensor.

Tests of our theory should be close at hand. Two recent experiments by Lan\c con  {\it et al.} \cite{lancon} and by Volpe {\it et al.} \cite{volpe} involve Brownian particles floating in inhomogeneous media. In that case all evidence suggests that the spatial inhomogeneity, due to hydrodynamic interactions of the fluid with walls, does not influence the steady state, according to a well-known theory with state-dependent diffusion and viscosity \cite{sanchodurr}. Similar experiments are conceivable where the medium is heated up in an nonuniform way, in which case our theory predicts the spatial density $p \propto T^{-1/2}$, a Maxwellian distribution of velocities near a given point, and prescribes how to shape a state space where uniform temperature Brownian motion occurs with the same statistical properties. Curvilinear Brownian motion could in principle be obtained by shaping thin recipients. This unfortunately requires us to deal with the friction of the particle with the walls in the direction of the relevant degree of freedom. If the steady-state profile would probably not be affected by this, all other statistical properties (including fluctuation theorems) would hardly be saved. Rather, at times when optical traps allow spectacular control of potentials confining Brownian particles \cite{grimm,ciliberto}, it might be conceivable to design one-dimensional optical curves. Another experimental signature of our theory might regard the occurrence of an anomaly between overdamped and finite-inertia thermodynamics, which in the work of Celani {\it et al.} is ascribed to the locally non-Gaussian nature of their steady state \cite{celani}, and might therefore be resolved in ours.
 
To conclude, we used a ``top-down'' approach to derive an ideal theory of diffusive systems in nonuniform temperature environments. As far as equilibrium steady states are concerned, the theory is coherent with several foundational aspects in equilibrium statistical mechanics, in particular the equivalence of statistical and physical entropies. Out of equilibrium, it entails a generalized version of the first law when the system's internal energy and the environmental temperature differ along a thermodynamic transformation. The theory allows us to describe nonequilibrium particle transport due to temperature drops and particle reservoirs, prescribing testable relationships between boundary values of temperatures, densities, and currents. The property of general covariance can be  constructively exploited to build a geometric analog. So far the environmental temperature profile was assumed to be given and stationary. Relaxing this hypothesis, it is tempting to advance a possible role of the theory with respect to Fourier's law of conduction, whose derivation is one major open question about systems with varying temperature \cite{bonetto}. ``Bottom-up'' microscopic derivations are very desirable, as, for example, in the spirit of Ref. \cite{balak}, with the purpose of establishing the margins of applicability of our theory and others.  Finally, nonequilibrium stochastic thermodynamics was only briefly touched on. References \cite{imparato,chetrite} discuss Langevin-type formalism, but their results cannot be directly applied to our theory, as they don't include state-dependent diffusion coefficients and Christoffel terms. A suitable generalization of fluctuation theorems, large deviation properties, and the full machinery of stochastic path integrals should and shall be applied to make precise connections between our several peculiar notions of entropy, heat and energy with measurable quantities.

\paragraph*{Aknowledgments.}

The author is grateful to the participants to the conference ``Computation of Transition Trajectories and Rare Events in Non-Equilibrium Systems'' held in Lyon, June 2012 --- among them S. Ruffo, H. Touchette, D. Lacoste, A. Patelli, U. Seifert, who discussed with the author in front of a last-minute poster. The research was supported by the National Research Fund Luxembourg in the frame of project FNR/A11/02.

\appendix

\section{Overdamping limit \label{overdamping}}

We derive the Fokker-Planck equation Eq.(\ref{eq:firstorder}) from the Kramers equation Eq.(\ref{eq:kramers}), in the limit of $\lambda \to \infty$. The treatment traces out Schuss's \cite[\S8.2]{schuss}. Solutions to the perturbative chain of equations are simply given, as they are known from previous literature. A logically  close step-by-step derivation can be operated by means of expansions in terms of Hermite polynomials, as is done in Ref.\cite{matsuo}.

Let $\epsilon = \gamma^{-1}$ be a small parameter. We rescale time according to $t \to \epsilon^{-1} s$ in the Kramers equation, and expand to order $\epsilon$ both the generator and the probability density:
\be
\Big(\epsilon^2 L_2 +  \epsilon L_1 + L_0 \Big) \Big(P_0 + \epsilon P_1+  \epsilon^2 P_2 + \ldots\Big) = 0,
\ee
where
\bes
L_0  & = &  \partial_v\left( v  + T \partial_v \right),  \\ 
L_1 & = & - v \partial - \partial_v \left( \frac{v^2}{2} \frac{\partial T}{T} + f_0  \right), \\
L_2 & = & - \partial_s.
\ees
We obtain the following expansion:
\bes
L_0 P_0 & = & 0,  \label{eq:L0} \\
L_0 P_1 + L_1 P_0 & = & 0,  \label{eq:L1} \\
L_0 P_2 + L_1 P_1 + L_2 P_0 & = & 0, \label{eq:L2} \\
& \ldots &   \nonumber
\ees
The solution to the first equation is
\be
P_0(x,v,s) = q_0(x,s) \exp - \frac{v^2}{2T(x)},
\ee
where $q_0(x,s)$ is an undetermined function. The corresponding spatial density is found by performing the integration over the velocities,
\be
p_0(x,s) = \int dv \, P_0(x,v,s) = N \sqrt{T(x)}\, q_0(x,s). \label{eq:suppspatial}
\ee
We also define the conditional probability
\be
\pi(v,s|x) = \frac{P_0(x,v,s)}{p_0(x,s)} = \frac{1}{N\sqrt{T(x)}}  \exp - \frac{v^2}{2T(x)} ,
\ee
which is a normal distribution in the velocity. Plugging  $P_0$ into Eq.(\ref{eq:L1}), we obtain
\be
\partial_v   \left( v  P_1 + T  \partial_v P_1  \right) =   v \left[ \partial  \ln (T  q_0 ) - f_0/T\right] P_0.
\ee
A solution is given by
\be
P_1 = \left\{ q_1  - v \left[ \partial \ln (T q_0) - f_0/T\right] \right\} P_0,
\ee
where $q_1= q_1(x,s)$ is again an undetermined function. Substituting into Eq.(\ref{eq:L2}) and integrating with respect to velocities gives
\be
 \frac{\partial p_0}{\partial s}= \frac{\partial}{\partial s}   \int P_0 dv 
= - \int v \, \partial P_1  dv, \label{eq:pass}
\ee
where it is assumed that all terms decay sufficiently fast at high velocities, so that
\bes
\int   \partial_v  \left[ \left( v  + T  \partial_v \right)  P_2 \right] dv  & = & 0, \\
\int  \partial_v \left[\left(- v^2 \partial T/2T \right)  P_1 \right] dv  & = & 0.
\ees
Equation (\ref{eq:pass}) then yields
\be
 \partial_s p_0 = - \int v \,  \partial \left\{ \left[ q_1  - v \, \partial \ln (Tq_0) + v f_0/T\right] P_0 \right\} dv.
 \ee
The first integrand is odd and its contribution vanishes. In view of Eq.(\ref{eq:suppspatial}), we obtain
\be
 \partial_s p_0 =  \partial \left\{[  \partial \ln (p_0\sqrt{T} ) - f_0/T]  \int v^2 P_0 \, dv  \right\}.
 \ee
 We now perform the integration
 \bea
 \int v^2 P_0(x,v,s) \, dv & = & p_0(x,s) \int v^2 \pi(v|x) dv \nonumber \\
 & = &  T(x) p_0(x,s),
 \eea
 where we recognized the covariance of $\pi(\cdot,x)$. Finally:
\bea
\frac{\partial p_0}{\partial s}  & = & \partial \left\{p_0 T  \left[  \partial \ln \left(\sqrt{T} p_0\right) - f_0/T\right] \right\} \\
& = & \partial \left\{ \left[ \sqrt{T}   \partial \left(\sqrt{T} p_0 \right) - p_0 f_0\right] \right\}
 \eea
which is Eq.(\ref{eq:firstorder}) to first order in $\epsilon$.

\end{document}